\documentclass[12pt]{article}
\newcommand{\STRUT}{\rule{0in}{2.45ex}}

\begin{document}
\begin{center}

{\large {\bf Phase Shifts and Mixing Parameters for Elastic Proton-Deuteron
      Scattering}}\\

\vspace*{0.80in}

A. Kievsky\\
Istituto Nazionale di Fisica Nucleare\\
Piazza Torricelli 2\\
56100 Pisa, Italy \\

\vspace*{0.20in}

J. L. Friar \\
Theoretical Division \\
Los Alamos National Laboratory \\
Los Alamos, NM  87545 \\

\vspace*{0.20in}

G. L. Payne\\
Department of Physics and Astronomy\\
University of Iowa\\
Iowa City, IA 52242\\

\vspace*{0.20in}

S. Rosati  \\
Dipartimento di Fisica\\
Universita' di Pisa \\
Piazza Torricelli 2\\
56100 Pisa, Italy \\

\vspace*{0.20in}

and

\vspace*{0.20in}

M. Viviani\\
Istituto Nazionale di Fisica Nucleare\\
Piazza Torricelli 2\\
56100 Pisa, Italy \\

\end{center}

\vspace*{0.50in}

\date{October 2000}

\begin{abstract}

The eigenphase shifts and mixing parameters for elastic p-d scattering below
the deuteron-breakup threshold are calculated for the AV14 two-body potential
using two different numerical methods. The excellent agreement 
confirms
that it is possible to perform accurate numerical studies of p-d scattering as
well as n-d scattering for low energies. The numerical results can be
considered as benchmarks for future calculations.

\end{abstract}
%\pacs{52.25.Vy, 94.10.Nh, 98.40.Bs}

\pagebreak

\centerline{\bf I. Introduction}
\medskip

The theoretical investigation of the three-nucleon system is an important tool
to test our knowledge of the nuclear interaction. This system is simple enough
that one can obtain accurate solutions for the $^3$H and $^3$He bound-state
systems. Modern realistic nucleon-nucleon two-body potential models that fit
the data with a $\chi^2$ per datum very close to unity underbind the trinucleon
bound states by 0.5 - 1.0 MeV. A possible solution to this discrepancy is to
include the three-nucleon interaction in the model Hamiltonian. Various models
of the three-nucleon force have been adjusted so that the correct three-body
binding energy is obtained. Since the other bound-state observables scale with
the binding energy, the bound-state system cannot be used to test our
understanding of the three-body force. 

Initially it was believed that the
nucleon-deuteron scattering problem could be used to test our understanding of
the three-nucleon force; however, it has been shown \cite{report} that most of
the scattering data can be reproduced at a good level with only two-body
interactions. For low-energy scattering the effects of the three-nucleon force
are usually small; there are nevertheless some discrepancies (such as the
nucleon analyzing power $A_y$ and the deuteron analyzing power $iT_{11}$) 
that could be sensitive to these interactions. Since
much of the experimental data is for p-d scattering at low energies, where the
Coulomb interaction cannot be neglected, it is necessary to solve the
scattering equations for this case as well as for the n-d case. 

Various groups have published benchmark calculations showing that it possible 
to obtain accurate solutions of the n-d Faddeev scattering equations for 
energies above the breakup threshold \cite{benmrk,benchmark}
and using $s$-wave potentials. Recently, a benchmark calculation for
n-d scattering has been published at $E_{lab}=3$ MeV (just below the
deuteron breakup threshold) using a realistic NN potential~\cite{nd-benchmark}.
Conversely, at present there has been no thorough study of the
accuracy of the solutions for p-d scattering. In this paper we present a
detailed comparison of the p-d phase shifts and mixing parameters obtained by
two different configuration-space calculations. These results can serve as
benchmarks for the low-energy p-d scattering problem. The methods used for
these calculations can be extended above the deuteron-breakup threshold, and
some preliminary calculations have been published for this case
\cite{pd-breakup}.

\medskip
\centerline{\bf II. Results}
\medskip

The two methods used for the results in this paper have been described in
previous papers. The Pisa group \cite{KVR93,pisa,K97} uses the
Pair-correlated Hyperspherical Harmonic (PHH) basis to expand the wave
function, and the corresponding S-matrix is obtained using the complex form of
the Kohn variational principle. The Los Alamos-Iowa (LA-Iowa) group 
\cite{chen,chapter} 
solves the Faddeev-Noyes equations using a spline expansion in
configuration space with boundary conditions appropriate to two-body Coulomb
scattering in the asymptotic region.

For the two-body potential we choose the Argonne AV14 model \cite{AV14}, since
this was used in a previous benchmark paper for n-d scattering
\cite{nd-benchmark}. We consider the same parameters and energies in order to
show the effects of the Coulomb interaction on the results. There is one small
difference from the previous calculations, 
we use $\hbar^2/M$~(MeV$\cdot$fm$^2$)$=41.47$ instead
of the value of $41.473$ used previously. 
The conventions used for the phase
shifts and mixing parameters are the same as in Ref.~\cite{nd-benchmark}. For
completeness, we show in Table 1 the binding energies and scattering lengths
for this potential. The Pisa values shown in the table were previously published
in Ref. \cite{K97}, while the LA-Iowa numbers are the results of recent more
accurate calculations. Also, we give in Table 2 the n-d results for our value
of $\hbar^2/M$. Our primary results, the p-d phase shifts and mixing
parameters, are given in Table 3, where one can see that the two calculations
differ by less than $1\%$, and in many cases (such as the large phase shifts)
the differences are much smaller than that.

\medskip
\centerline{\bf III. Conclusions}
\medskip

The results from two very different numerical methods are in excellent
agreement for p-d scattering at energies below the deuteron-breakup threshold.
This confirms that in this case it is possible in practice to perform
theoretical studies with the same precision as for the n-d problem. These
results also provide a benchmark against which researchers in the future can
test their techniques.

\medskip
\centerline{\bf Acknowledgments}
\medskip

The work of J.\ L.\ F.\ was performed under the auspices of the U.\ S.\
Department of Energy. That of G.\ L.\ P.\ was supported in part by the U.\ S.\
Department of Energy.

\pagebreak

\begin{table}
\begin{center}
\begin{tabular} {|c|cc|}
\hline
& Pisa & LA-Iowa \\
\hline
$E_B (^3{\rm H})$  & 7.684 & 7.684 \\
$E_B (^3{\rm He})$ & 7.033 & 7.033 \\
\hline
$^2a_{\rm nd}$ & 1.189 & 1.191 \\
$^4a_{\rm nd}$ & 6.379 & 6.376 \\
$^2a_{\rm pd}$ & 0.941 & 0.945 \\
$^4a_{\rm pd}$ & 13.773 & 13.752 \\
\hline

\end{tabular}
\end{center}
\caption{\label{tab01} Comparison of the binding energies in MeV and scattering 
lengths in fermis determined by the Kohn variational method (Pisa) and the
configuration-space Faddeev equations (LA-Iowa) for the AV14 NN potential.}

\end{table}

\pagebreak

\begin{table}
\vspace*{-0.75in}
\begin{center} 
\begin{tabular} {|c|c||r|r||r|r||r|r||}
\hline
\multicolumn{2}{|c||}{} & \multicolumn{2}{c||}{$E_{\rm lab}$ = 
  1 MeV} & \multicolumn{2}{c||} {$E_{\rm lab}$ = 2 MeV} & 
  \multicolumn{2}{c||}{$E_{\rm lab}$ = 3 MeV} \\
\hline

\multicolumn{1}{|c|}{$J^{\pi}$}&\multicolumn{1}{c||}{$\delta_{\Sigma\lambda}$}
&\multicolumn{1}{|c|}{Pisa}&\multicolumn{1}{|c||}{LA-Iowa}
&\multicolumn{1}{|c|}{Pisa}&\multicolumn{1}{|c||}{LA-Iowa}
&\multicolumn{1}{|c|}{Pisa}&\multicolumn{1}{|c||}{LA-Iowa}\\

\hline
    &$\delta_{\frac{3}{2}2}$&-0.999&-0.999&-2.573&-2.573&-3.903&-3.901\\
${1\over2}^{+}$				   
    &$\delta_{\frac{1}{2}0}$&-17.696&-17.702&-27.866&-27.864&-34.822&-34.812\\
    &$\eta$           & 1.043& 1.041& 1.205& 1.203& 1.253& 1.251\\
\hline                                     
    &$\delta_{\frac{1}{2}1}$&-4.195&-4.195&-6.654&-6.651&-7.532&-7.526\\
${1\over2}^{-}$				   
    &$\delta_{\frac{3}{2}1}$& 12.384& 12.383& 20.558& 20.551& 25.052& 25.026\\
    &$\epsilon$       & 3.742& 3.742& 5.394& 5.393& 7.256& 7.255\\
\hline                                     
    &$\delta_{\frac{3}{2}0}$&-47.148&-47.143&-61.329&-61.334&-70.491&-70.532\\
    &$\delta_{\frac{1}{2}2}$& 0.580& 0.580& 1.547& 1.547& 2.419& 2.418\\
${3\over2}^{+}$				   
    &$\delta_{\frac{3}{2}2}$&-1.073&-1.073&-2.771&-2.770&-4.214&-4.212\\
    &\STRUT$\epsilon$       & 0.651&0.650&0.716&0.716&0.780&0.780\\
    &\STRUT$\xi$            & 0.545&0.546& 1.010& 1.010& 1.438& 1.437\\
    &\STRUT$\eta$           &-0.114& -0.114& -0.247& -0.247& -0.387& -0.388\\
\hline                                     
    &$\delta_{\frac{3}{2}3}$& 0.125& 0.125& 0.502& 0.503& 0.943& 0.943\\
    &$\delta_{\frac{1}{2}1}$&-4.139&-4.139&-6.490&-6.486&-7.197&-7.188\\
${3\over2}^{-}$				   
    &$\delta_{\frac{3}{2}1}$& 14.285& 14.287& 22.722& 22.721& 26.398& 26.382\\
    &$\epsilon$       &-1.307&-1.309&-1.970&-1.972&-2.761&-2.766\\
    &\STRUT$\xi$      &-0.184&-0.185&-0.269&-0.269&-0.258&0.258\\
    &\STRUT$\eta$     &-1.108&-1.110&-2.316&-2.318&-3.804&-3.807\\
\hline                                           
    &$\delta_{\frac{3}{2}4}$&-0.015&-0.015&-0.093&-0.093&-0.211&-0.211\\
    &$\delta_{\frac{1}{2}2}$& 0.575& 0.575& 1.528& 1.529& 2.384& 2.384\\
${5\over2}^{+}$				    
    &$\delta_{\frac{3}{2}2}$&-1.141&-1.141&-2.974&-2.973&-4.567&-4.564\\
    &$\epsilon$             &-0.288&-0.288&-0.309&-0.308&-0.328&-0.327\\
    &\STRUT$\xi$            &-0.287&-0.287&-0.522&-0.521&-0.737&-0.735\\
    &\STRUT$\eta$           &-0.870&-0.870&-1.573&-1.572&-2.156&-2.154\\
\hline                                          
    &$\delta_{\frac{3}{2}1}$& 13.442& 13.452& 22.026& 22.029& 26.347& 26.350\\
    &$\delta_{\frac{1}{2}3}$&-0.065&-0.065&-0.257&-0.257&-0.476&-0.476\\
${5\over2}^{-}$				    
    &$\delta_{\frac{3}{2}3}$& 0.131& 0.131& 0.523& 0.524& 0.971& 0.971\\
    &$\epsilon$             &-0.468& -0.468& 0.493& 0.493& 0.517& 0.512\\
    &\STRUT$\xi$            & 0.414& 0.415& 0.728& 0.729& 0.985& 0.985\\
    &\STRUT$\eta$           &-0.131&-0.131&-0.255&-0.255&-0.361&-0.361\\
\hline

\end{tabular} \end{center}
\caption{\label{tab02} Comparison of the n-d phase shifts and mixing parameters
(in degrees) determined by the Kohn variational method (Pisa) and the
configuration-space Faddeev equations (LA-Iowa) for the AV14 NN potential.}

\end{table}

\pagebreak

\begin{table}
\vspace*{-0.75in}
\begin{center} 
\begin{tabular} {|c|c||r|r||r|r||r|r||}
\hline
\multicolumn{2}{|c||}{} & \multicolumn{2}{c||}{$E_{\rm lab}$ = 
  1 MeV} & \multicolumn{2}{c||} {$E_{\rm lab}$ = 2 MeV} & 
  \multicolumn{2}{c||}{$E_{\rm lab}$ = 3 MeV} \\
\hline

\multicolumn{1}{|c|}{$J^{\pi}$}&\multicolumn{1}{c||}{$\delta_{\Sigma\lambda}$}
&\multicolumn{1}{|c|}{Pisa}&\multicolumn{1}{|c||}{LA-Iowa}
&\multicolumn{1}{|c|}{Pisa}&\multicolumn{1}{|c||}{LA-Iowa}
&\multicolumn{1}{|c|}{Pisa}&\multicolumn{1}{|c||}{LA-Iowa}\\

\hline
    &$\delta_{\frac{3}{2}2}$&-0.787&-0.787&-2.281&-2.281&-3.615&-3.615\\
${1\over2}^{+}$				   
    &$\delta_{\frac{1}{2}0}$&-12.607&-12.616&-23.525&-23.535&-31.400&-31.414\\
    &$\eta$           & 1.189& 1.187& 1.257& 1.256& 1.259& 1.260\\
\hline                                     
    &$\delta_{\frac{1}{2}1}$&-3.394&-3.393&-6.122&-6.120&-7.405&-7.401\\
${1\over2}^{-}$				   
    &$\delta_{\frac{3}{2}1}$& 9.431& 9.432& 17.896& 17.894& 22.815& 22.799\\
    &$\epsilon$       & 3.128& 3.133& 4.589& 4.593& 6.227& 6.229\\
\hline                                     
    &$\delta_{\frac{3}{2}0}$&-37.335&-37.323&-53.436&-53.436&-63.673&-63.705\\
    &$\delta_{\frac{1}{2}2}$& 0.454& 0.454& 1.357& 1.356& 2.206& 2.204\\
${3\over2}^{+}$				   
    &$\delta_{\frac{3}{2}2}$&-0.849&-0.849&-2.459&-2.459&-3.905&-3.903\\
    &\STRUT$\epsilon$       &0.819&0.817&0.792&0.796&0.833&0.836\\
    &\STRUT$\xi$            & 0.533&0.532& 0.976& 0.974& 1.385& 1.384\\
    &\STRUT$\eta$           & -0.095& -0.093& -0.215& -0.213& -0.343& -0.340\\
\hline                                     
    &$\delta_{\frac{3}{2}3}$& 0.100& 0.100& 0.447& 0.447& 0.872& 0.873\\
    &$\delta_{\frac{1}{2}1}$&-3.358&-3.358&-6.007&-6.005&-7.166&-7.158\\
${3\over2}^{-}$				   
    &$\delta_{\frac{3}{2}1}$& 10.957& 10.958& 20.152& 20.150& 24.628& 24.615\\
    &$\epsilon$       &-1.077&-1.080&-1.650&-1.653&-2.327&-2.333\\
    &\STRUT$\xi$            &-0.191&-0.190&-0.296&-0.295&-0.322&0.321\\
    &\STRUT$\eta$           &-0.998&-0.994&-2.091&-2.092&-3.362&-3.365\\
\hline                                           
    &$\delta_{\frac{3}{2}4}$&-0.011&-0.011&-0.081&-0.081&-0.194&-0.193\\
    &$\delta_{\frac{1}{2}2}$& 0.451& 0.451& 1.342& 1.342& 2.176& 2.175\\
${5\over2}^{+}$				    
    &$\delta_{\frac{3}{2}2}$&-0.905&-0.905&-2.640&-2.639&-4.229&-4.227\\
    &$\epsilon$       &-0.378&-0.378&-0.353&-0.355&-0.362&-0.364\\
    &\STRUT$\xi$            &-0.318&-0.320&-0.530&-0.529&-0.736&-0.734\\
    &\STRUT$\eta$           &-0.997&-1.006&-1.617&-1.616&-2.189&-2.186\\
\hline                                          
    &$\delta_{\frac{3}{2}1}$& 10.237& 10.246& 19.279& 19.289& 24.216& 24.222\\
    &$\delta_{\frac{1}{2}3}$&-0.051&-0.051&-0.229&-0.228&-0.443&-0.442\\
${5\over2}^{-}$				    
    &$\delta_{\frac{3}{2}3}$& 0.103& 0.103& 0.464& 0.464& 0.899& 0.899\\
    &$\epsilon$       & -0.234& -0.223& 0.259& 0.261& 0.360& 0.361\\
    &\STRUT$\xi$            & 0.409& 0.408& 0.729& 0.729& 0.992& 0.992\\
    &\STRUT$\eta$           &-0.136&-0.135&-0.258&-0.257&-0.366&-0.365\\
\hline
\end{tabular} \end{center}
\caption{\label{tab03} Comparison of the p-d phase shifts and mixing parameters
(in degrees) determined by the Kohn variational method (Pisa) and the
configuration-space Faddeev equations (LA-Iowa) for the AV14 NN potential.}

\end{table}

\end{document}